\documentstyle[preprint,aps,psfig,epsf]{revtex}
\newcommand{\lan}{\langle}
\newcommand{\ran}{\rangle}
\begin{document}
\title{Dynamical response functions in models of vibrated granular media}
\author{Mario Nicodemi}
\pagestyle{myheadings}
\address{Dipartimento di Fisica, Universit\'a di Napoli ``Federico II'',
Unit\`a INFM and INFN Napoli \\
Mostra d'Oltremare, Pad. 19, 80125 Napoli, Italy}

\maketitle
\date{\today}
\begin{abstract}
In recently introduced schematic lattice gas models for 
vibrated dry granular media, we study the dynamical response of the system 
to small perturbations of shaking amplitudes 
and its relations with the characteristic fluctuations. 
Strong off equilibrium features appear and 
a generalized version of the fluctuation dissipation theorem is 
introduced. The relations with thermal glassy systems and the role of 
Edwards' compactivity are discussed. 
\end{abstract}
\vskip2pc


Few years ago, Edwards formulated the hypothesis 
that it is possible to extend to powders the methods 
of standard statistical mechanics \cite{Edwards}, an important challenge 
since powders are by definition ``non thermal" systems \cite{JNBHM}. 
Actually, in granular media the role of a ``temperature", linked to the 
concept of Edwards' compactivity \cite{Edwards}, is played by the amplitude 
of external vibrations \cite{JNBHM}, and it is possible to properly 
individuate the corresponding 
``equilibrium" states \cite{NC_aging,EdwardsGrinev,Novak}. 
However, these systems are typically in off equilibrium configurations, 
as the presence of aging phenomena shows \cite{NC_aging,Head}, and the above 
extension of statistical mechanics thus runs across its off equilibrium 
version. 

A very important issue for both theoretical and practical reasons, 
is the understanding of relations between the response of a system 
to external perturbations and its characteristic fluctuations. 
Here, we study such a problem in the context of schematic lattice gas models 
\cite{CH,NCH,Caglioti} recently introduced to describe gently shaken 
granular systems. We show how in granular media 
it is possible to formulate a fluctuation dissipation theorem
(FDT) where the shaking amplitudes play the role of usual temperatures, 
but in typical off equilibrium situations it coincides neither 
with its usual version at equilibrium nor with the extensions 
valid in off equilibrium thermal systems in the so called ``small 
entropy production" limit \cite{BCKM}. 
We also discuss the important relations with Edwards' theory. 


The lattice gas model we consider here, the Tetris, schematically describes 
the effects of steric hindrance and geometric frustration on grains 
in granular materials. Interestingly it shows many phenomena typical of 
granular media under shaking, as logarithmic compaction or segregation 
\cite{NCH,Caglioti,NC_aging}. 
It consists of a system of elongated grains, with two possible orientations, 
which occupy the sites of a square lattice tilted by $45^{\circ}$. 
To avoid overlaps, two particles can be nearest 
neighbor only if they have the right reciprocal orientation.
The motion of grains in absence of ``vibrations" is subject only to gravity 
so grains only move downwards (without overlapping). 
The effect of vibration is introduced by allowing the particles to diffuse 
upwards with a probability $p_{up}$ and downwards with $p_{down}=1-p_{up}$.
The effects of geometric frustration on the microscopic dynamics 
are introduced with a kinetic constraint: particles can flip their 
orientation only if at least three of their neighbors are empty.

The parameter governing the dynamics is the adimensional quantity 
$x_0=p_{up}/p_{down}$, or $\Gamma\equiv 1/\ln(x_0^{-1/2})$. 
$\Gamma$ plays the same role as the amplitude of the vibrations 
in real granular matter \cite{JNBHM}, 
and, for not too high amplitudes, 
a good agreement is found between the model and experiments by 
posing $\Gamma^b\sim a/g$ with $b$ about 1 
(here $g$ is the gravity and $a$ the peak acceleration of the shakes 
in the experiments) \cite{NCH,NC_aging}. 
In analogy with experiments, in our Monte Carlo simulations the shaking 
of the system is as follows: we prepare the system, confined in a box, 
in a given initial configuration at $t=0$ (see below), 
then we start to ``shake" it continuously and indefinitely with a 
given ``amplitude" $x_0$. We expect very similar results 
by considering, instead of a single long tap, a series of short taps, as
experimentally more convenient (see Ref.~\cite{Novak,Knight}).

\begin{figure}[ht]
\centerline{\psfig{figure=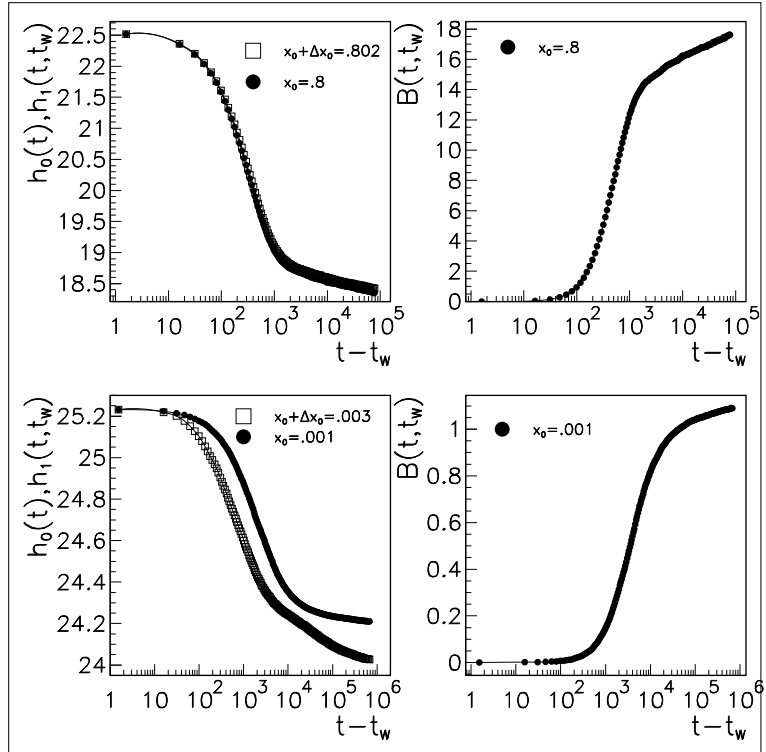,height=10cm,width=10cm,angle=0}}
\caption{
{\em Top left} The average grains height $h_0(t)$ (filled circles) and 
$h_1(t,t_w)$ (empty squares) as a function of time $t-t_w$, 
of a lattice granular system initially prepared in a 
uniform fluidized state and then shaken at $x_0=0.8$, and of a replica 
perturbed after a time $t_w=370$ by shaking at $x_1=x_0+\Delta x_0=0.802$. 
{\em Top right} The ``displacement'', 
$B(t,t_w)\equiv C_0(t,t)-2C_0(t,t_w)+C_0(t_w,t_w)$ 
(where $C_0(t,t')=\lan h_0(t)h_0(t')\ran $), of the above system at $x_0=0.8$  
as a function of $t-t_w$. 
{\em Bottom left} The average height $h_0(t)$ (filled circles) and 
$h_1(t,t_w)$ (empty squares) as a function of $t-t_w$, of a system 
initially prepared in a static compact state 
and then shaken at $x_0=0.001$, and of a replica 
shaken at $x_1=.003$ after a $t_w=333$. 
{\em Bottom right} The ``displacement'', 
$B(t,t_w)$, of the system at $x_0=0.001$ as a function of $t-t_w$. 
} 
\label{ut8}
\end{figure}

The Tetris can be described with the following Ising Hamiltonian  
in the limit $J\rightarrow\infty$ (see \cite{NCH,Caglioti}): 
$\beta H=J\sum_{\langle ij\rangle } f_{ij}(S_i,S_j)n_i n_j 
- \frac{g}{T}\sum_i y_i$ . 
Here $g$ is the gravity constant, $y_i$ is the $i$-th particle height, 
$n_i=0,1$ are occupancy variables, $S_i=\pm 1$ are Ising spin variables 
for the two possible orientations of grains. 
$f_{ij}$ is a function describing the hard core repulsion ($J=\infty$):
$f_{ij}(S_i,S_j) = 0$ if the orientation of neighbors, $(S_i,S_j)$, 
is allowed and $f_{ij}=1$ if it is not allowed 
(
$f_{ij}(S_i,S_j)=1/2\left[S_i S_j -\epsilon_{ij} (S_i+ S_j) +1\right]$, where 
$\epsilon_{ij}=+ 1 $ for bonds along one direction of the lattice and 
$\epsilon_{ij}=- 1 $ for bonds along the other). 
The temperature, $T$, of the above Hamiltonian system  
is related to the ratio $x_0=p_{up}/p_{down}$ via $e^{-2g/T} = x_0$
(i.e., $\Gamma=T/g$).


This Hamiltonian mapping shows that in our model the field coupled to 
grains height is the adimensional gravity: 
$g/T=\ln(x_0^{-1/2})$. Thus a perturbation to 
the system, coupled to a easy observable, may be introduced by varying $x_0$. 
We record, with Monte Carlo simulations, 
the dynamical correlation functions and the response of the system 
to such a small perturbation in the ``shaking amplitude''. 

Inside a box of fixed size $30 \times 60$, with periodic boundary conditions 
along the x-axis and rigid walls at bottom and top, we prepare the system  
in a uniform density initial configuration ($\rho=0.5$), corresponding 
to an highly fluidized state, and then we ``shake'' it at a given 
amplitude $x_0$. 
During this process we record the average height 
$h_0(t)=\lan \sum_iy_i(t)\ran$ of the grains and the two times correlation 
function $C_0(t,t')\equiv \lan h_0(t)h_0(t') \ran$, or, the 
``mean square displacement" $B(t,t')\equiv C_0(t,t)-2C_0(t,t')+C_0(t',t')$ 
which is a quantity relevant to test the FDT theorem.
Time is measured in such a way that unity corresponds to an 
average update of all the degrees of freedom in the system, and 
the statistical averages run over 2048 noise and initial configuration 
realizations.
In order to measure the response function, 
we also record the average height of grains, $h_1(t,t_w)$, in 
an identical copy of the system (a ``replica''), which, 
after a fixed time $t_w$ (typically below we fix $t_w=370, 3700$), 
is shaken at $x_1=x_0+\Delta x_0$, i.e., is perturbed by a small increase of 
the shaking amplitude $\Delta x_0$ 
(in what follows $\Delta x_0=0.002$ \cite{nota1}). 

The quantity $\Delta h(t,t_w)=h_1(t,t_w)-h_0(t)$, i.e., the difference 
in heights between the perturbed and unperturbed systems, 
is by definition the integrated response. 
FDT or its generalizations concern the relation between 
$\Delta h(t,t_w)$ and the displacement, $B(t,t_w)$, which is linked to the 
correlations in the unperturbed system. 
The simplest version of a possible generalization 
to off equilibrium granular matter of the FDT 
may be argued from thermal systems \cite{BCKM,CuLeD}. 
The integrated response, $\Delta h$, should be approximately 
proportional to $B$:
\begin{equation}
\Delta h(t,t_w) \simeq \frac{X}{2} \Delta(\Gamma^{-1}) B(t,t_w)
\label{FDT}
\end{equation}
Here, $\Delta(\Gamma^{-1})\equiv \Gamma_0^{-1}-\Gamma_1^{-1}$, 
is the variation between the inverse shaking amplitudes of the reference and 
the perturbed systems and the prefactor, $X$, is a quantity to be determined, 
in principle function of $t_w$ and $t$ themselves. 
In thermal system, in the limit $t,t_w\rightarrow\infty$, $X$ is a 
piecewise constant, depending just on $B(t,t_w)$ and not on both $t$ 
and $t_w$ \cite{BCKM,CuLeD}.

In the specific case of our model we have 
$\Delta(\Gamma^{-1})\equiv \Delta(\frac{g}{T})= \ln\left[(
\frac{x_0+\Delta x_0}{x_0})^{-\frac{1}{2}}\right]$, and 
if $X=1$ we recover the usual well known equilibrium version of FDT. 
In the study of glassy thermal systems the quantity $\Gamma/gX\equiv T/X$ 
has the meaning of an ``effective temperature" of the sample, 
which only for $X=1$ coincide with the equilibrium bath temperature 
\cite{BCKM}. 

\begin{figure}[ht]
\centerline{\psfig{figure=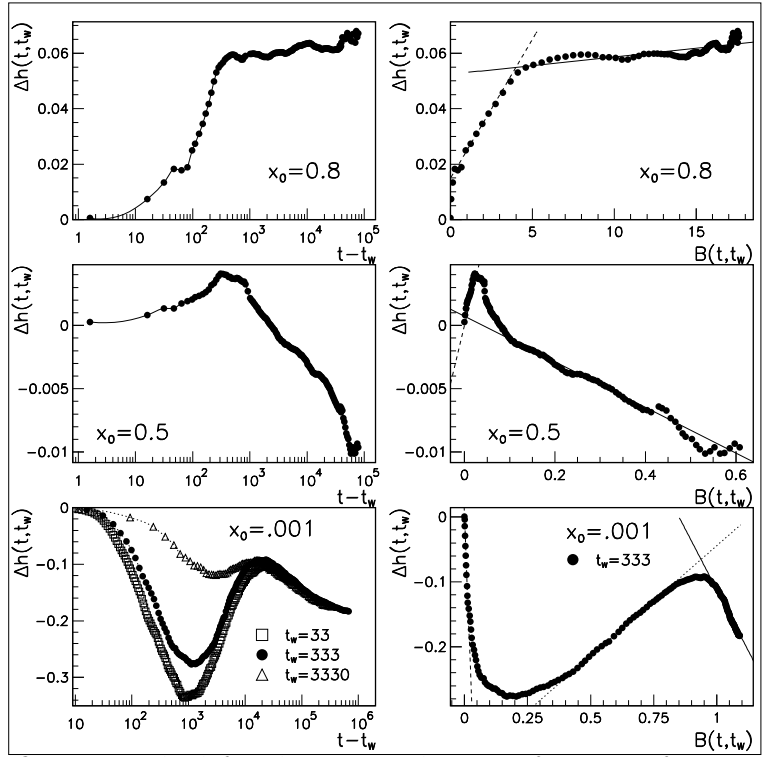,height=10cm,width=10cm,angle=0}}
\caption{
In the figures on the left column, 
we plot, as a function of $t-t_w$, the average height difference, 
$\Delta h(t,t_w)\equiv h_1(t,t_w)-h_0(t)$, 
of a reference system shaken at a given $x_0$ and a replica perturbed 
after $t_w$ by shaking at $x_0+\Delta x_0$ ($\Delta x_0=.002$). 
In the right column, in order to check the generalized 
fluctuation dissipation theorem (FDT) of eq.~(\ref{FDT}), 
we plot the quantity, $\Delta h(t,t_w)$, i.e., the integrated response, 
as a function of the displacement of the reference system, $B(t,t_w)$. 
The top and middle cases correspond to systems initially prepared in 
a uniform fluidized state, then shaken at different ``amplitudes'' $x_0$ 
($x_0=0.8$ top, $x_0=0.5$ middle), with replicas perturbed after a $t_w=370$. 
In the bottom figure, the systems, initially prepared in 
a static compact state, is 
shaken at $x_0=0.001$ and its replica is perturbed 
by $\Delta x_0=0.002$ after a $t_w=33, 333, 3330$. 
As an ``hot'' gas at equilibrium has a higher average height respect 
to a ``colder'' gas, $\Delta h$ should be always positive. 
However, at low $x_0$ ($x_0=0.5, 0.001$), negative responses appear.
In agreement with eq.~(\ref{FDT}), $\Delta h$ is asymptotically still 
approximately piecewise linear in $B$, but, up to the time scales we 
explored, only at $x_0=0.8$ the equilibrium version of FDT holds. 
The superimposed linear curves indicate the different regimes described 
in the text. 
}            
\label{ut_all}
\end{figure}

In the present model, eq.~(\ref{FDT}) seems to be approximately valid, 
also if in typical off-equilibrium situations $X$ slowly depends on $t$ 
and $t_w$.
To outline this, as first we explore the high shaking amplitude regime. 
In Fig.\ref{ut8} (top left), as a function of $t-t_w$ are shown, 
for $t_w=370$ (analogous results are found for $t_w=3700$), 
the height $h_0(t)$ of a system ``shaken" at $x_0=0.8$, and $h_1(t,t_w)$ 
of the replica at $x_1=x_0+\Delta x_0=0.802$. $h_0$ and $h_1$ 
decrease in time signaling that the system is slowly compactifying
while approaching the equilibrium density profile.
In Fig.\ref{ut8} (top right), is also shown the 
``displacement", $B(t,t_w)$, of the same unperturbed system.  
We show in Fig.\ref{ut_all} (top left) the difference 
$\Delta h(t,t_w)\equiv h_1(t,t_w)-h_0(t)$ as a function of $t-t_w$. 
In agreement with the expected simple behavior of a gas subject to gravity 
(which reacts to an increase of temperature, or decrease of gravity,
by increasing its average height), $\Delta h$ is a positive quantity:
the ``colder" system (with $x_0=0.8$)
actually must have a lower equilibrium height respect the ``hotter" one
(with $x_1=0.8+\Delta x_0$). 
In order to check the generalization of the FDT proposed 
in eq.~(\ref{FDT}), in Fig.\ref{ut_all} (top right), 
we plot $\Delta h$ as a function of the displacement $B$. 
Actually, $\Delta h$ is approximately piecewise linear in $B$: 
after an early transient with $X>1$, in the 
long time regime $X$ is independent on $t$ and $t_w$ and equal to 1, 
showing that, in the high ``temperature" and 
low density region, the usual equilibrium version of FDT is obeyed. 

In the low $x_0$ region, we already know the presence of strong 
off equilibrium phenomena \cite{NC_aging}. Thus it is reasonable to expect 
that the above simple picture drastically changes. 
In Fig.\ref{ut_all} (middle left), we plot the integrated response,
$\Delta h(t,t_w)$, as a function of $t-t_w$ for a system ``shaken" at 
$x_0=0.5$ and a replica shaken at $x_0+\Delta x_0=.502$ after 
a $t_w=370$ (analogous results are found for $t_w=3700$). 
It is apparent that after an early transient, up to the time scales 
we explored we find a {\em negative response}, $\Delta h$, 
in complete contrast with the above equilibrium scenario. 
Actually, both replicas start from a strongly off equilibrium state, but 
while the ``colder" is ``frozen" and remains longly trapped in metastable 
states, the ``hotter" is able to more rapidly escape to approach its 
asymptote. 

Notice that $\Delta h$ is negative 
even after five decades in time and no change in its trend is observed. 
The approach to equilibrium is actually extremely slow
\cite{NCH,Caglioti}: if the region with 
negative response would extend up to infinity (which is hard to say with 
computer simulations), this should results in a 
dynamical breaking of ergodicity introduced by gravity in 
our system of particles interacting just with excluded volume effects. 
Interestingly these results are partially supported by some recent 
experiments \cite{Knight}, which showed that the asymptotic density 
of a granular system compactified at a low shaking amplitude from  
a random initial configuration, is lower than the analog quantity in a system 
shaken at a slightly higher amplitude. This result is in contrast with the 
``equilibrium" measures (the ``reversible branch") in the experiments of 
Novak et al. \cite{Novak}.
The above phenomenon has strong repercussions on the FDT. 
As plotted in Fig.\ref{ut_all} (middle right), 
$\Delta h$ may be approximately plotted 
as a piecewise linear function in $B$, as from eq.~(\ref{FDT}), 
but after an initial transient with $X>1$, the system enters a region with 
negative $X$ ($X\simeq -5$), corresponding to a negative 
``effective temperature" $T/X$. 
The specific value of $X$, in this region, also slowly depends on $t_w$ 
(see below), and we are far from the small entropy production limit known 
in thermal systems where $X=X(B)$. 

A similarly anomalous picture is found in a system shaken at 
$x_0=0.001$ and a replica at $x_0+\Delta x_0=.003$.
To show that the above general results do not depend on the 
details of the initial state, we discus the latter Monte Carlo 
shaking experiment in a differently prepared system, closer to experiments. 
Now the starting particle configuration is 
prepared by randomly inserting particles 
into the box from its top and then letting them fall down, with the
above dynamics, until the box is half filled. Thus, the system 
starts from a more compact static state. 
In Fig.\ref{ut8}, the heights of the original system, $h_0$, of the 
replica $h_1$ (bottom left), along with the displacement 
$B(t,t_w)$ (bottom right) are plotted as a function of $t-t_w$ for $t_w=333$. 
This figure shows again a negative response $\Delta h$ 
which is plotted as a function of $t-t_w$ in Fig.\ref{ut_all} (bottom left), 
over six order of magnitudes, 
for three different values of $t_w$ ($t_w=33, 333, 3330$). 
Three different regions are observed for $\Delta h$, but the short times 
transient inflection zone seems to disappear for $t_w$ long enough. 
In agreement with the generalized FDT of eq.~(\ref{FDT}), 
$\Delta h$ is at long times ($10^4 \leq t-t_w\leq 10^6$) again approximately 
linear in $B$, as shown for $t_w=333$ in Fig.\ref{ut_all}
(bottom right) \cite{nota2}. The negative proportionality 
coefficient, $X$, slowly depends on $t_w$ 
(from $X\sim -7$ for $t_w=33$ to $X\sim -15$ for $t_w=3330$). 
For systems prepared 
in the previous fluidized initial state, discussed above, 
analogous results are found. 
It is important to stress that the above behaviors with negative responses 
are not found if the system starts from an equilibrium initial configuration. 

Interestingly, in the high density or low $x_0$ region, we observe an 
analogous scenario also in a similar lattice model, 
the IFLG \cite{NCH}, where, in order to describe the motion of grains in a 
disordered environment, quenched disorder is introduced in the Hamiltonian
described above, but no kinetic constraints are present in the dynamics.

As stated, the presence of negative $X$ in the generalized FDT, is a 
feature of the very far from equilibrium dynamics. 
Actually, this is not expected for instance in off equilibrium 
thermal systems as glasses or spin glasses, at least in the small entropy 
production limit \cite{BCKM}. Moreover, in spin glasses the parameter $X$, 
which is a seemingly dynamical quantity, may be asymptotically related to 
equilibrium static properties from replica theory \cite{BCKM,CuKu}. 
In granular media, as long as in fragile glasses \cite{ParisiMD}, 
the static correspondent of $X$, if any, is still missing. 
It is useful to underline that while in glasses an experimental measure 
of the generalized FDT may be non trivial, in granular media, 
as shown above, this should be reasonably simple. 

The interesting finding in our models for granular matter 
of a negative $X$, corresponding to negative responses and 
``effective temperatures", may be clarified by stressing 
the relations with Edwards' theory \cite{Edwards}.

Actually, the present approach to granular media based on a 
standard Hamiltonian formalism, is very close in spirit to the more 
general statistical mechanics approach proposed by Edwards \cite{Edwards}. 
The hard core repulsion term in the above Hamiltonian $H$ 
has a role similar to  Edwards' volume function $W$ (where a constraint 
of mechanical stability is explicitly present). The fundamental control 
parameter in Edwards' theory which corresponds to our adimensional 
temperature, $T/g$, is the compactivity $\lambda X_E$. Edwards and 
Grinev \cite{EdwardsGrinev} guess that the compactivity is related to the 
experimental ``temperature" of a shaken granular medium according to the 
following ``fluctuation-dissipation" relation: $\lambda X_E \sim (a/g)^2$ 
(as above, $a$ is the shake and $g$ is the gravity acceleration). 
This statement is the analog of our suggestion, $(T/g)^b \sim a/g$, 
which is based on the comparison of Monte Carlo and experimental data. 

In Ref.\cite{EdwardsGrinev} is proposed that the state of  
a shaken granular medium in the equilibrium regime, corresponding 
for instance to configurations on the 
so called ``reversible branch" in the experiments of Ref.~\cite{Novak}, 
is characterized by a positive compactivity, and, by extrapolation, 
the off equilibrium dynamics, corresponding 
to the experimental ``irreversible branch" of Ref.~\cite{Novak}, 
has a negative compactivity, $X_E<0$.
This observation is in agreement with our discover of different 
regions with 
positive as long as negative effective temperatures, $T/X$, 
in the study of the generalized FDT in granular media. 
A negative off equilibrium compactivity may correspond to the necessity 
of the system to cross states with ``higher entropy" in order to 
lower its volume under shaking. 
Actually, the measure of integrated responses v.s. correlation functions 
in granular materials may open the way to a direct experimental access to 
Edwards' compactivity for a clear settlement of a statistical mechanics 
for such systems. 


In summary, in order to understand the off equilibrium statistical mechanics 
of powders, we have studied, in schematic lattice gas models for vibrated 
dry granular media, the 
dynamical response functions to small shaking amplitude 
perturbations and their relations to characteristic dynamical fluctuations. 
Strong off equilibrium features appeared, 
as long time regions with negative response functions,
different from those observed in the small entropy 
production limit of thermal glassy systems \cite{BCKM}, 
along with the necessity to 
introduce a generalized version of the fluctuation dissipation theorem. 
The novel properties we have found, as the presence of negative effective 
temperatures, are consistent with Edwards' theory 
of powders and demand important experimental check. 

\smallskip

We thank INFM-CINECA for computer time on Cray-T3D/E.
Work partially supported by the TMR Network Contract ERBFMRXCT980183 and 
MURST-PRIN 97.


\begin{thebibliography}{40}
\bibitem{nota1} $\Delta x_0$ is small to ensure the 
linear response regime.
\bibitem{nota2}
Here the ratio $\Delta x_0/x_0$ is no longer 
small, and thus the linear response regime might not hold.
\end{thebibliography}
\end{document}